# An Efficient Deep Learning Image Condition for Locating Earthquakes

*Wenhuan Kuang[1,2], Jie Zhang[4], and Wei Zhang[2,3]\**

[1]*Department of Marine Geosciences, Key Lab of Submarine Geosciences and Prospecting Techniques, Ocean University of China, Qingdao 266100, China*

[2]*Department of Earth and Space Sciences, Southern University of Science and Technology, Shenzhen, 518055, China*

[3]*Guangdong Provincial Key Laboratory of Geophysical High-resolution Imaging Technology, Southern University of Science and Technology, Shenzhen 518055, China.*

[4]*Department of Geophysics, University of Science and Technology of China, Hefei, 230026, China*

*Corresponding author: Wei Zhang(zhangwei@sustech.edu.cn)*

**Abstract**

**Migration-based earthquake location methods may encounter the polarity reversal issue due to the non-explosive components of seismic source, leading to an unfocused migration image. Various methods have been proposed, yet producing an ideally focused migration source image is still a challenge. In this study, by taking advantage of the general pattern recognition ability of convolutional neural network, we propose a novel Deep Learning Image Condition (DLIC) to address this issue. The proposed DLIC measures the goodness of waveform alignments for both P- and S-waves after correcting their traveltime moveouts and it follows the geophysical principle of seismic imaging that the best-aligned waveforms utterly represent a best-imaged source location. Both a synthetic test and real data application are used to show the effectiveness and merits of the proposed DLIC. A test on synthetic data shows that the DLIC can effectively overcome the polarity reversal issues in the data. Real data application to southern California shows that the DLIC can greatly enhance the focusing of migrated source image over the widely used source scanning algorithm. Further tests show that the DLIC is applicable to continuous seismic data, to regions with few historical earthquakes, and has the potential to locate small earthquakes. The proposed DLIC shall benefit the migration-based source location methods.**

## I. INTRODUCTION

Earthquake location is one of the most fundamental tasks in seismology. Conventionally, methods for locating earthquakes can be classified into two main categories: traveltime-based methods and migration-based methods. Traveltime-based methods [1][2] rely on the accurate picking of P- and S-phases [3][4][5] as well as robust phase association algorithms [6][7]. Relative location methods such as the master event method [8] and the double-difference method [9][10] can also be classified as traveltime-based methods since they utilize the relative traveltime differences resolved from either first arrival picking or waveform cross-correlation. Template matching [11][12][13] utilizes waveform similarity and emerges to be an effective tool in revealing unprecedented details of seismicity if an abundant catalog of template events is available. Migration-based earthquake location methods [14], such as back-projection [15][16] and diffraction stacking [17][18][19], have the merit of locating weak events when their first arrivals are difficult to pick. Migration-based methods usually require a proper image condition to anticipate a well-focused source image. However, due to the non-explosive source radiation pattern of natural or microseismic earthquakes, the waveform polarities may be reversed at different recording azimuths (Fig. 1a). This is known as the polarity reversal issue [20][21][22], which usually leads to a quasi-symmetric migrated source image (Fig. 1b).

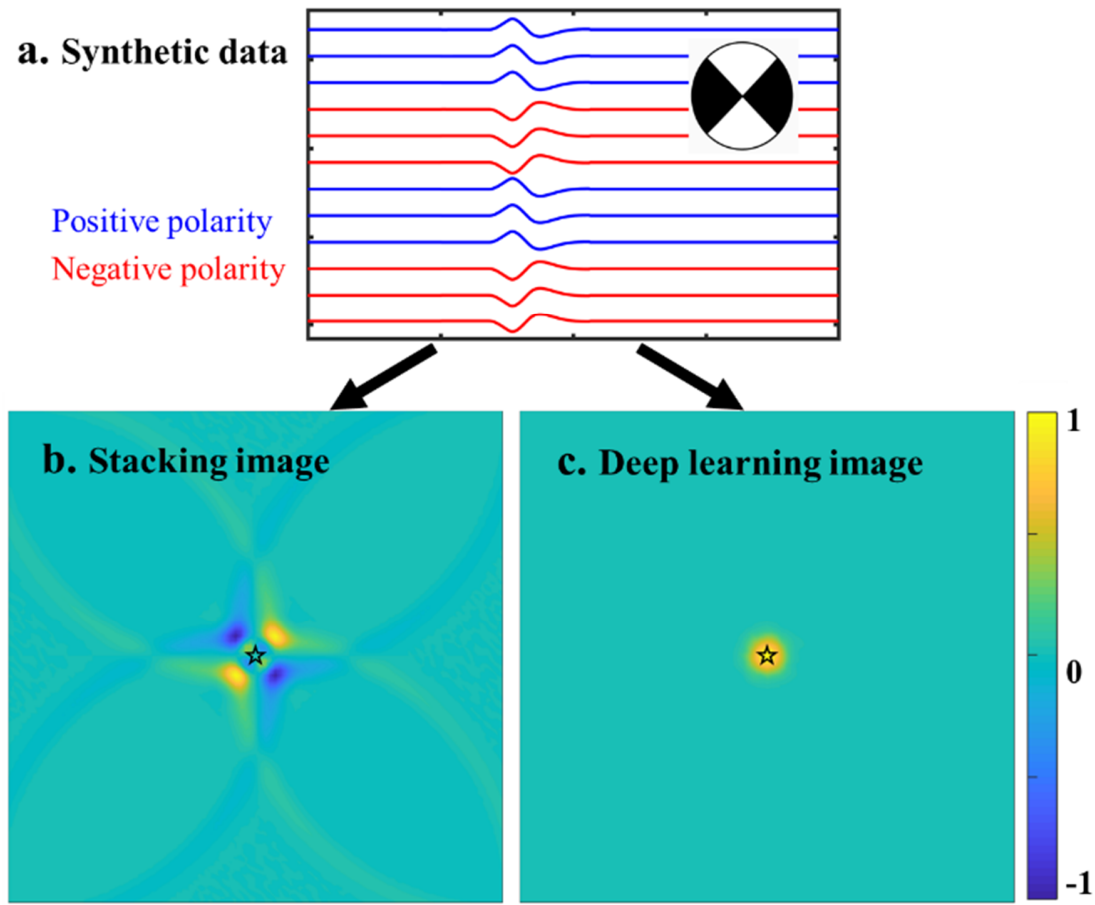

**Fig. 1. a** Synthetic waveforms generated with a strike-slip double couple source. Blue and red waveforms denote positive and negative polarities, respectively. **b** Migrated image derived using diffraction stacking [18]. **c** Deep-learning image derived using the proposed DLIC neural network. The black star denotes the true location.

With such an unfocused source image, it is challenging to correctly retrieve the event location because the maximum value is not centered at the ground-truth event location (black star in Fig. 1b). It is desirable to produce an ideally migrated source image (Fig. 1c) with a proper image condition. Previous studies have proposed to calibrate the polarity inconsistency before migration [21] or to jointly invert the source focal mechanism along with location [23][24][25], yet the computational costs in such solutions are very expensive. To address this issue, our previous study [22] has proposed a two-step procedure, in which we first perform conventional diffraction migration [18] and then transform the migrated image into a Gaussian probability distribution using a neural network. In this study, we propose an entirely new Deep Learning Image Condition (DLIC) that can be directly applied to the observed waveforms and thus greatly simplifies the procedure. The new DLIC is demonstrated to be very effective in solving the polarity reversal issue as well as enhancing the focusing of the migration image.

A test on the synthetic data (Fig. 1a) which is modelled from a pure double couple focal mechanism shows that the proposed DLIC can produce an ideally focused deep-learning image (Fig. 1c). The maximum image value in the predicted deep-learning image is centered at the ground-truth event location (black star in Fig. 1c), which greatly outperforms the conventional stacking image derived from the diffraction migration (Fig. 1b). To better illustrate this new method, we organize the study as follows: first, we illustrate the methodology; then, we illustrate the data augmentation and training process; after that, we apply the method to a real data set in southern California to show its effectiveness and followed by discussions and conclusions.

## II. Method and Network Architecture

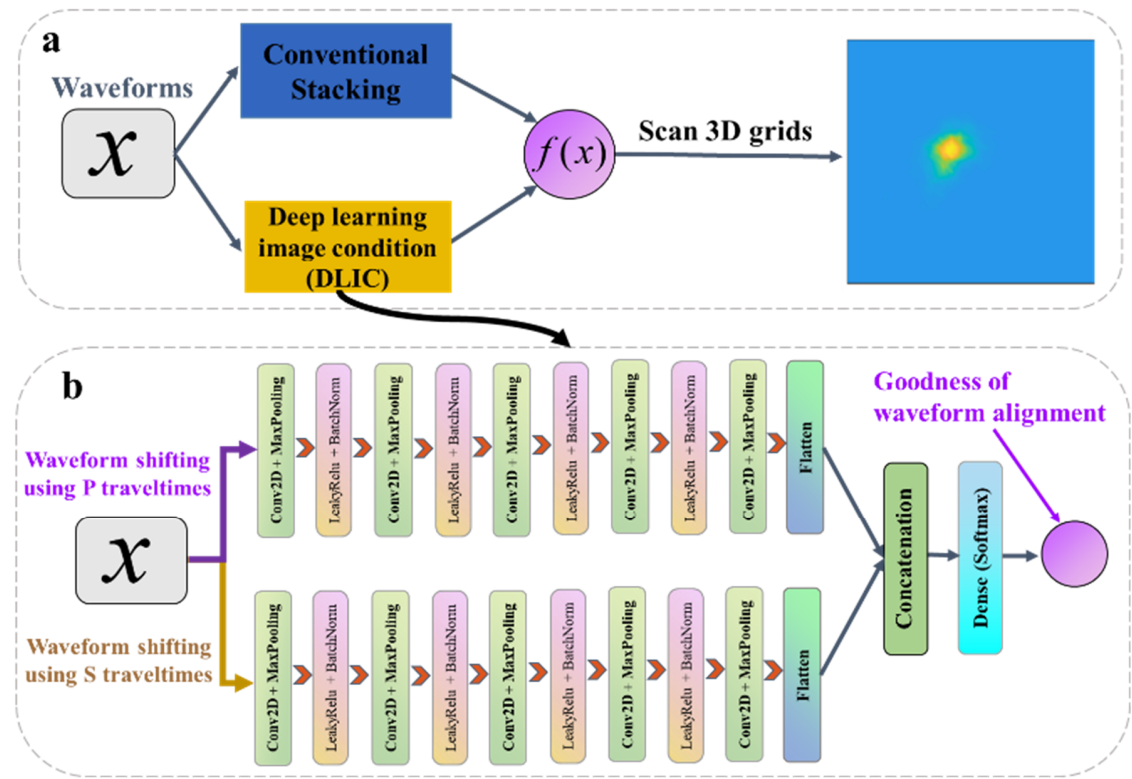

**Fig. 2. a** Illustrating the proposed DLIC by analogy with conventional stacking image condition. **b** The network architecture of DLIC.

The DLIC is analogous to the conventional diffraction stacking image condition (left part in Fig. 2a), where for each trial location, it inputs the observed waveforms and outputs an image value. The difference is that conventional diffraction stacking sums the waveform amplitudes along the theoretically calculated traveltime moveouts, but the DLIC uses a neural network to measure the goodness of waveform alignments for both the P- and S-waves after correcting their traveltime moveouts. In other words, the conventional diffraction stacking image condition measures the stacking energy and the DLIC measures the goodness of waveform alignments for both P- and S-waves. In the sense of seismic imaging [26][27][28], both the maximized stacking energy and the well aligned waveforms for both P- and S-waves equally represent an ideally imaged source. Therefore, the proposed DLIC follows the basic geophysical principle of seismic imaging.

The procedure of using the DLIC is thus similar to the conventional diffraction stacking method: first, we discretize the study area into three-dimensional (3D) spatial grids; then, we use the DLIC to predict the image values at all these spatial grids; finally, we can produce a deep-learning image (right panel in Fig. 2a) that is almost identical to conventional migration image. The predicted deep-learning image physically represents the goodness of waveform alignments for both the P- and S-waves by using each spatial grid to align the recorded waveforms. Subsequently, we can locate the earthquake by finding the maximum value in the deep-learning image.

The network architecture of the DLIC is shown in Fig. 2b. We mainly use convolutional, maxpooling, and dense layers as well as some necessary operations such as leakyrelu activation, batch normalization, and concatenation. In general, the DLIC network contains two subnetworks that first separately process the traveltime-corrected waveforms of P- and S-waves and then concatenates them to output a single value that measures the goodness of overall waveform alignments. Traveltime correction is realized by shifting the whole waveforms using either P- or S-wave theoretical traveltimes and we do not need to separate P- and S-waves. The traveltime tables of both P- and S-waves are calculated with the widely used 3D velocity model in southern California [33] in advance.

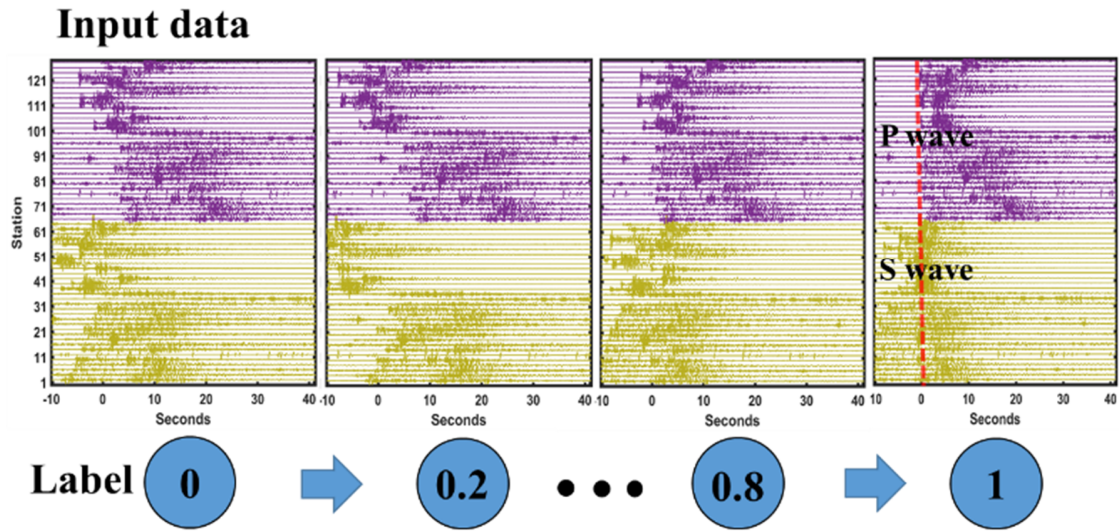

**Fig. 3.** Training input data and their corresponding training labels.

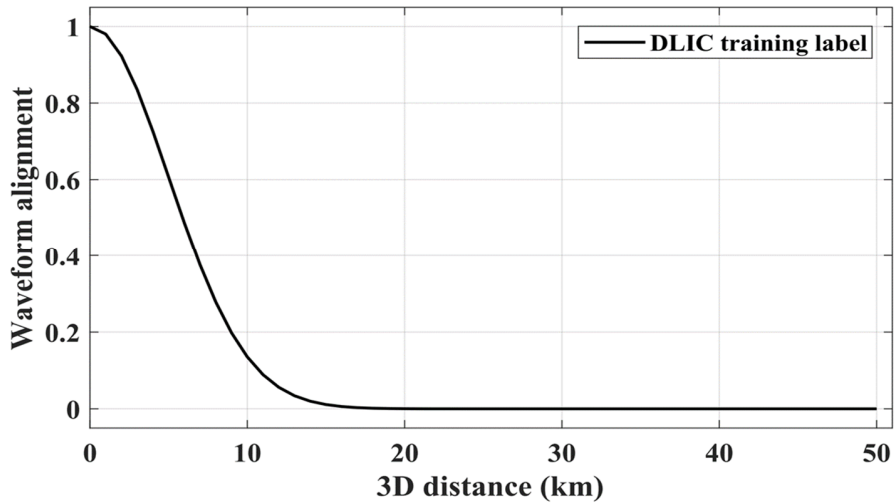

**Fig. 4.** Definition of training label based on 3D distance.

We define the training labels according to the alignments of both P- and S-waves after traveltime correction (Fig. 3). Considering the ground-truth locations utterly correspond to the best-aligned P- and S-waves, we mathematically define the training labels using the 3D distance between the virtual location and the ground-truth location (Fig. 4). The training labels follow the Gaussian distribution concerning and they range from 0 to 1, where small values represent poor waveform alignments and 1 represents the best goodness of waveform alignment (red dashed line in Fig. 3).

## III. Data Augmentation and Training

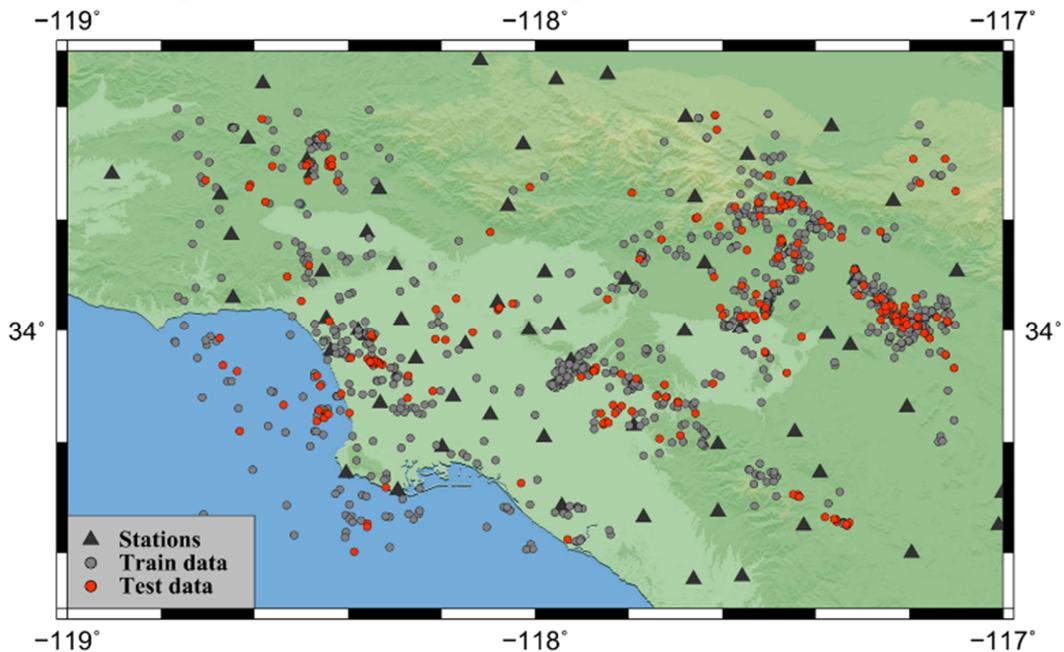

**Fig. 5.** Data distribution. Black triangles, grey dots, and red dots denote seismic stations, training data, and test data, respectively.

We download the seismic data of vertical components in southern California at 64 local stations from 2010 to 2021 with magnitudes larger than 2.0. We split the earthquakes that occurred from January 2010 to June 2019 as training data (grey dots in Fig. 5) and the earthquakes that occurred from July 2019 to September 2021 as test data (red dots in Fig. 5). Considering some cataloged earthquakes may be mislocated, it is necessary to clean the training data to better train the neural network. To do this, we conduct a manual check and remove those false earthquakes (falsely triggered noise or inaccurately located events) from the catalog. Overall, this data set contains 1036 earthquakes for training and 212 earthquakes for testing. For preprocessing, we apply a bandpass filter of 1 Hz to 5 Hz, a resampling rate of 0.05 seconds, and an amplitude normalization to the waveforms. After that, we cut the waveforms with a window length of 51.2 seconds.

To train the DLIC network, we adopt a data augmentation strategy to generate a diversity of traveltime-corrected waveforms based on those 1,036 training earthquakes. For each training earthquake, we first randomly generate 400 virtual locations around the ground-truth location (Fig. 6). The probability of generating these virtual locations decreases with distance to balance the training samples. Then, we use all these virtual locations to perform traveltime corrections on the original waveforms. Thus, we generate a significant number of augmented training samples (1,023×400) containing various kinds of traveltime-corrected waveforms.

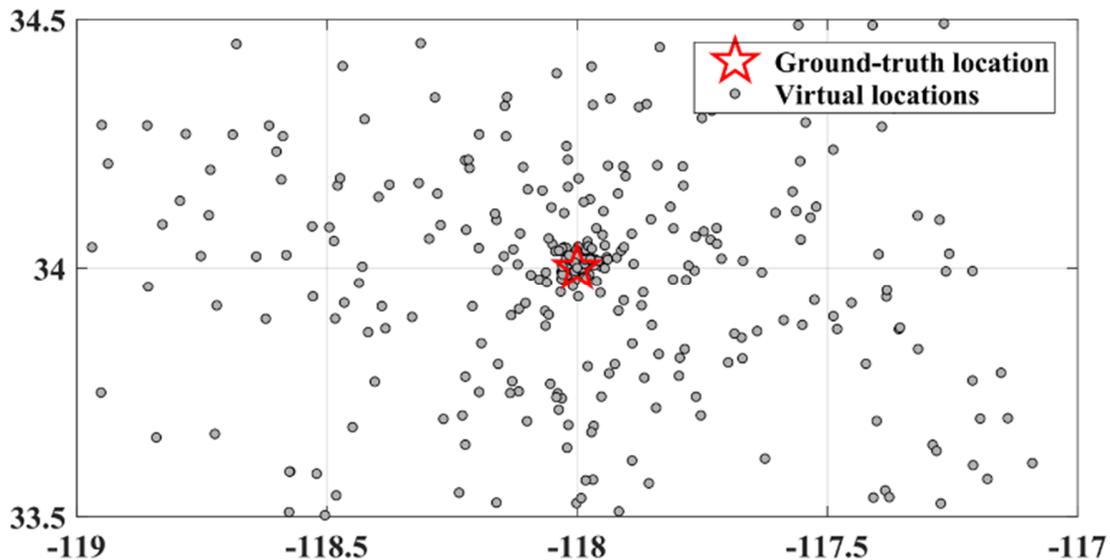

**Fig. 6.** Data augmentation strategy. Red star and grey dots denote the ground-truth location and augmented virtual locations, respectively.

## IV. Test Results in Southern California

Once the DLIC neural network is well trained, we test its performance on 212 unseen earthquakes (red dots in Fig. 5). First, we discretize the 3D spatial grids with the interval of 2 km in all three directions (i.e., longitude, latitude, and depth). Then, we loop all these virtual grids to predict their DLIC values. Finally, we can produce a deep-learning image for each event. Fig. 7 shows several examples of the predicted deep-learning images. Most of the predicted deep-learning images show very good focusing. Neural networks is demonstrated to have the property of recognizing the overall waveform characteristics [29][30][31][32] in the data and as demonstrated from the test results, the proposed DLIC is not susceptible to waveform polarity reversal issue. We can retrieve the optimal event location by finding the maximum image value in the deep-learning image.

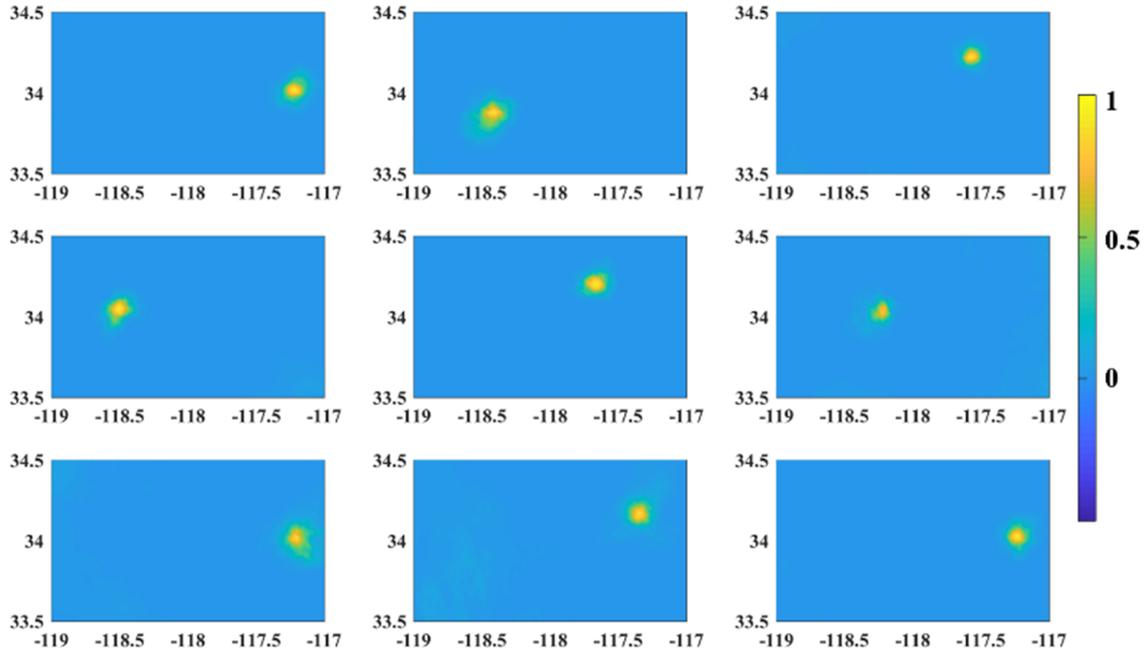

**Fig. 7.** Several examples of the predicted deep-learning images.

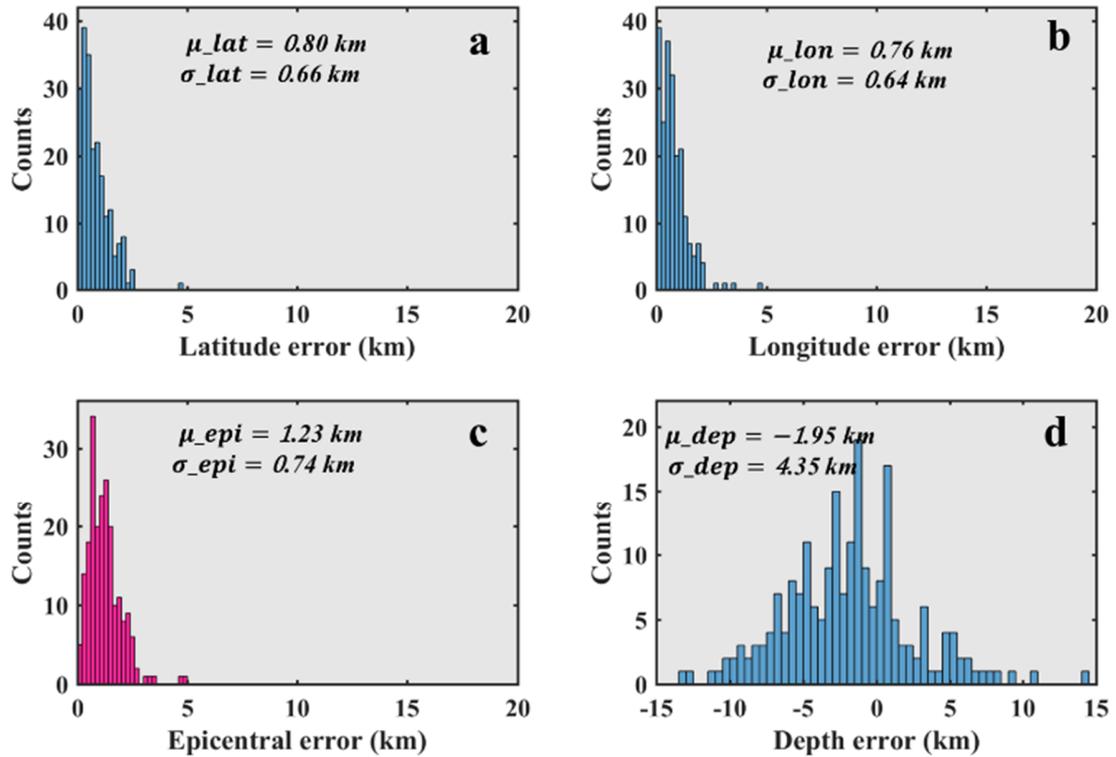

**Fig. 8.** Statistical analysis of the test results. **a** Latitude error. **b** longitude error. **c** Epicentral error. **d** Depth error.

The statistical analysis of the test results shows that the mean absolute errors in latitude (Fig. 8a), longitude (Fig. 8b), and epicentral distance (Fig. 8c) are 0.8 km, 0.76 km, and 1.23 km, respectively. The standard deviations in latitude, longitude, and epicentral distance are 0.66 km, 0.64 km, and 0.74 km, respectively. The mean error and standard deviation in depth (Fig. 8d) are -1.95 km and 4.35 km, respectively. Considering the uncertainties of the cataloged locations in southern California may reach 2 km [34], these test results demonstrate that the well-trained DLIC neural network can effectively locate the earthquakes with acceptable uncertainties.

To compare with the widely used Source Scanning Algorithm (SSA) [17], we derive both the deep-learning images (Fig. 9a and 9b) and the SSA stacking images (Fig. 9c and 9d) on the same earthquake data. We find that the 80% percentile and 60% percentile contours (dashed red curves in Fig. 9) of the deep-learning images are much narrower than that of the SSA stacking images, suggesting that the predicted deep-learning images exhibit much better energy concentration than the SSA method. This is because the DLIC particularly emphasizes recognition of those linear data patterns of best-aligned P- and S-waves and is thus less susceptible to other interference phases. Conversely, the SSA method may be blurred by interference phases such as scattering waves, multiples, and coda waves in the stacking process.

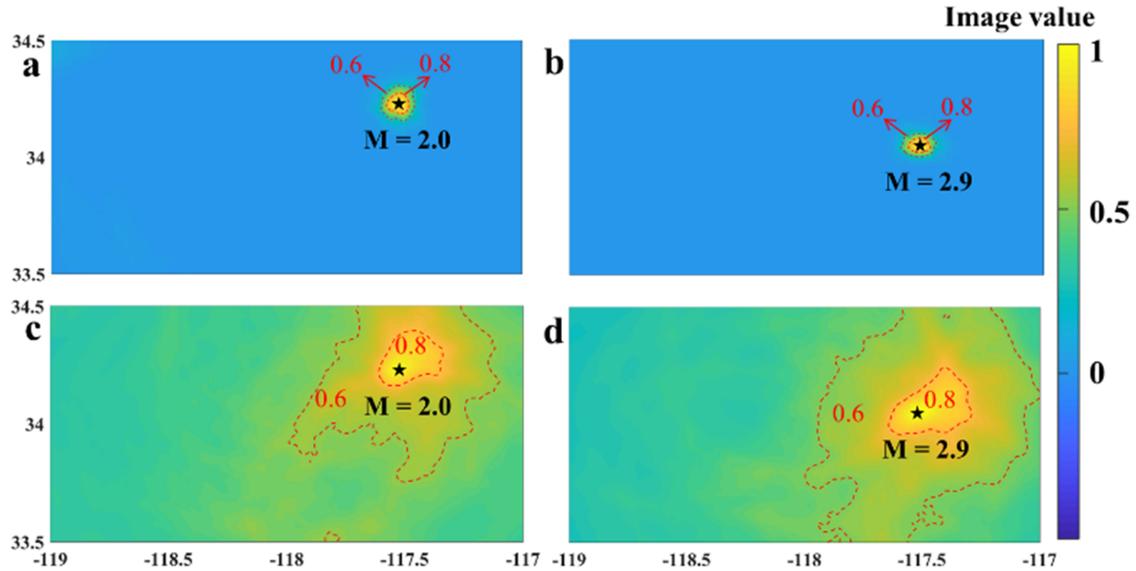

**Fig. 9.** Comparison between the deep-learning images and the conventional stacking images. **a** and **b** The deep-learning images for two events. **c** and **d** The conventional stacking images for the same events using the source scanning algorithm [17]. Dashed red contours denote values of 0.6 and 0.8. Black stars denote the ground-truth locations. Stacking images are normalized by their maximum value.

## V. Discussions

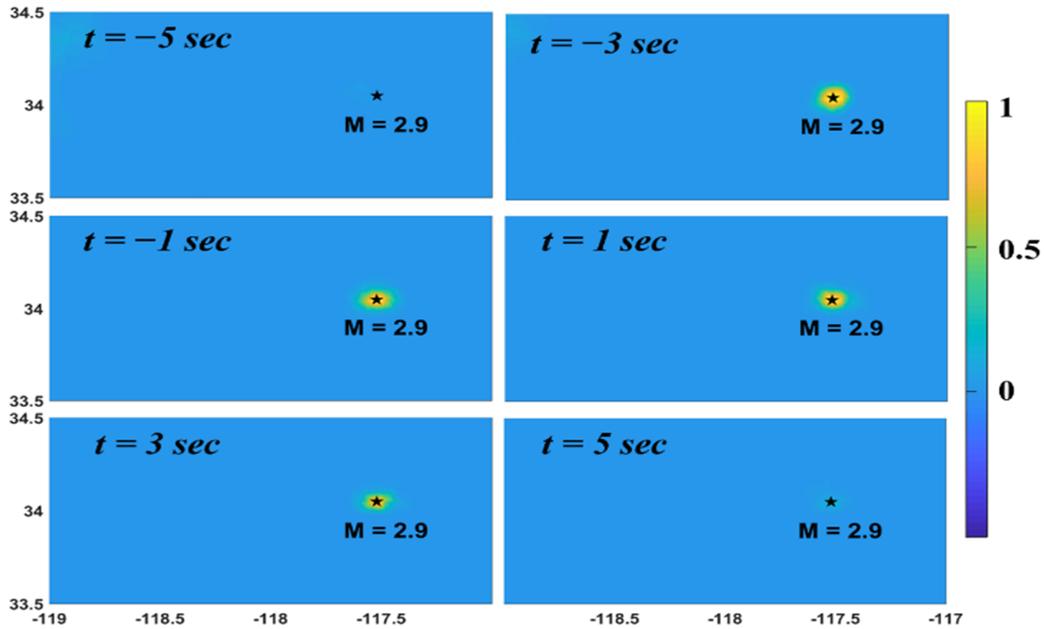

**Fig. 10.** Test on continuous seismic data. The black star denotes the ground-truth location.

The proposed DLIC can be applied to continuous seismic data. We test the DLIC on continuous seismic data and predict the deep-learning images every 2 seconds (Fig. 10). Test results show that we can retrieve the origin time with a decent resolution (error ≤2.0 seconds). The retrieved optimal event location from the deep-learning image is very close to the ground-truth location (black star in Fig. 10) with errors of 1.02 km and 0.03 km in latitude and longitude, respectively.

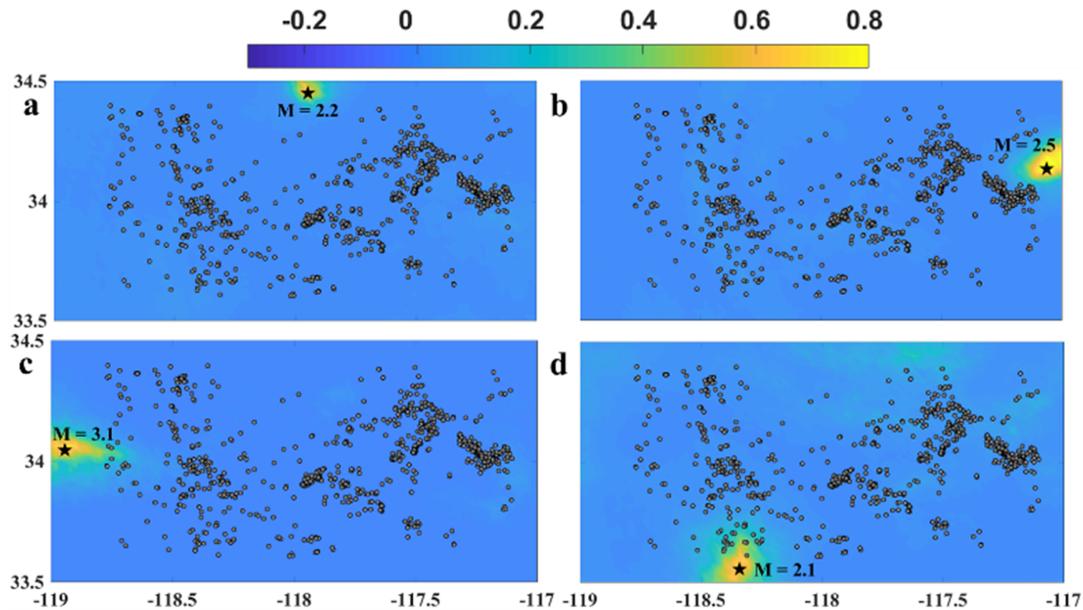

**Fig. 11.** Test on earthquakes that have no nearby historical earthquakes included for training. Grey dots denote earthquakes used for training. Black stars denote the ground-truth locations.

The DLIC can locate earthquakes that occur in regions with few historical earthquakes included for training. To validate such a property, we test four events (black stars in Fig. 11) that occurred in different regions where no nearby historical earthquakes are available for training. The predicted deep-learning images (Fig. 11) of the four events show decent energy focusing. The absolute epicentral errors of these four events are 1.37 km, 0.84km, 2.67 km, and 0.73 km, suggesting that the DLIC is generally applicable to regions where few historical earthquakes are available for training. This is because the DLIC is trained to recognize the linear data patterns of the best-aligned P- and S-waves. No matter where a new earthquake may occur, the linear data patterns of the best-aligned P- and S-waves will eventually emerge while scanning the 3D virtual grids. But the focus of the predicted deep-learning images may slightly degrade.

Furthermore, the proposed DLIC has the potential to locate small earthquakes. We test 50 small events with 1≤M<2, although the DLIC network is trained on events with M≥2. Test results (Fig. 12) show that 80% of epicentral errors are within 10 km. If we have an accurate location catalog for small events, we can retrain the DLIC and achieve much better performance. Nevertheless, it indicates that the proposed DLIC has the potential to locate small earthquakes.

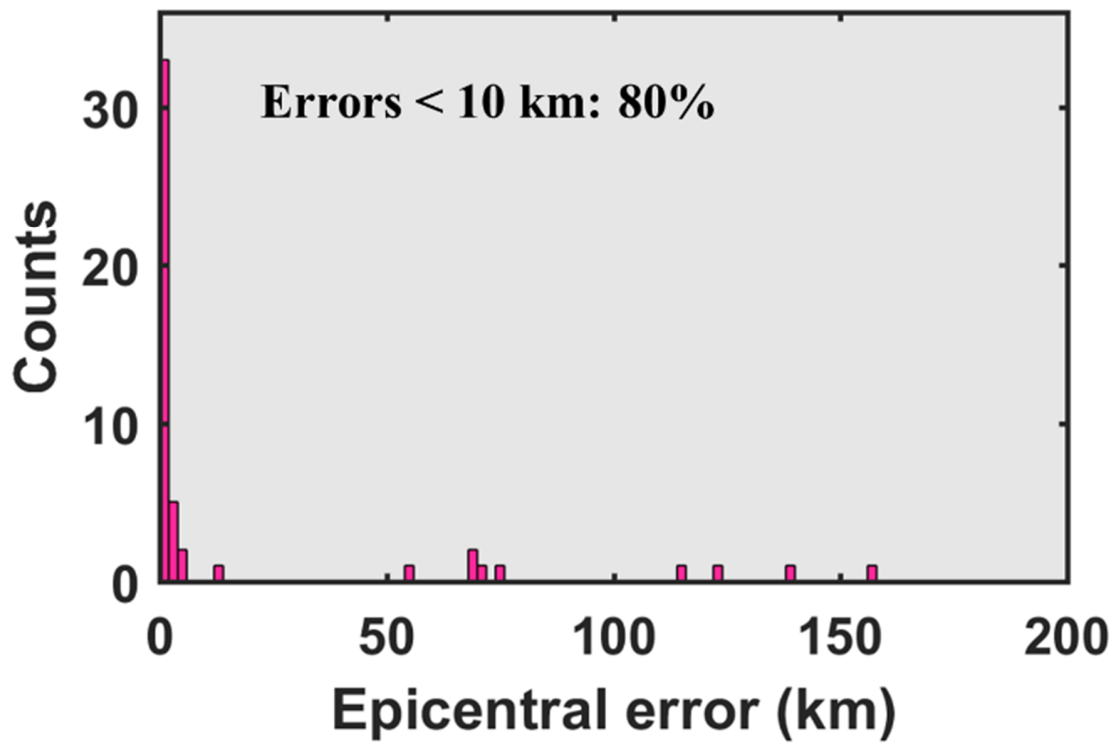

**Fig. 12.** Test on smaller earthquakes with 1≤M<2. Note that the DLIC network is trained on earthquakes with M≥2.

# VI. Conclusion

In this study, we have proposed a novel DLIC for locating earthquakes. Both a synthetic test and real data application are used to demonstrate the validity and effectiveness of the DLIC. The synthetic test shows that the proposed DLIC can produce an ideally focused source image. Real data application demonstrates that the DLIC can generate a better-focused deep-learning image than the conventional stacking image. In discussions, we show that the DLIC can be applied to continuous seismic data and to regions where few historical earthquakes are available for training. Moreover, we show that the proposed DLIC has the potential to locate small earthquakes. In general, the proposed DLIC is very effective in solving the polarity reversal issue and enhancing the focusing of the migrated image and it shall benefit migration-based location methods.

ACKNOWLEDGMENT

We thank the financial support from the National Key R&D Program of China (Grant No. 2021YFC3000703-05), the National Natural Science Foundation of China (Grant No. 42104047, U1901602), and the Guangdong Provincial Key Laboratory of Geophysical High-resolution Imaging Technology (2022B1212010002). We thank Dr. Congcong Yuan at Harvard University for helpful discussions.